\documentclass[twocolumn, a4paper, 10pt, nofootinbib]{revtex4}
\usepackage[top=20mm,bottom=20mm,left=15mm,right=15mm]{geometry}
\usepackage{amsmath, amssymb, amsfonts, mathtools, comment, cases, braket, amsthm, bm, mathdots, cancel, bm}

\numberwithin{equation}{section} 

\usepackage{hyperref}
\hypersetup{
setpagesize=false,
 bookmarksnumbered=true,%
 bookmarksopen=true,%
 colorlinks=true,%
 linkcolor=blue,
 citecolor=red,
}

\newcommand{\be}{\begin{equation}}
\newcommand{\ee}{\end{equation}}
\newcommand{\ba}{\begin{eqnarray}}
\newcommand{\ea}{\end{eqnarray}}

\begin{document}

\title{Vainshtein screening in Horndeski theories nonminimally and kinetically coupled to ordinary matter}
\author{Kitaro Taniguchi and Ryotaro Kase}
\affiliation{
Department of Physics, Faculty of Science, 
Tokyo University of Science, 1-3, Kagurazaka,
Shinjuku-ku, Tokyo 162-8601, Japan}

\begin{abstract}
We study the Vainshtein screening mechanism in Horndeski theories 
in the presence of a scalar field $\phi$ nonminimally and kinetically 
coupled to ordinary matter field. A general interacting Lagrangian describing 
this coupling is characterized by energy transfer $f_1$ and momentum exchange $f_2$. 
For a spherically symmetric configurations on top of the cosmological background, 
we investigate the static perturbations in linear and nonlinear regimes with respect to 
the scalar field perturbation. 
In the former regime, the parametrized post-Newtonian parameter generally deviates from unity  
as long as the matter coupling or $G_{4,\phi}$ exists. On the other hand, 
in the latter regime, we show that the nonlinear self-interaction term of 
scalar field successfully activates the Vainshtein mechanism even 
in the presence of the couplings $f_1$ and $f_2$. The gravitational 
potentials recover the Newtonian behavior deep inside the Vainshtein radius. 
The bounds on coupling terms not to substantially change the Vainshtein 
radius are also given.

\end{abstract}

\date{\today}

\maketitle

\section{Introduction}
The late-time cosmic acceleration is discovered from the observation of 
Type Ia supernovae in 1998 \cite{riess1998Observational, perlmutter1999Measurements}. 
The unknown source of this accelerated expansion is named dark energy (DE). 
The simplest candidate for DE is the cosmological constant $\Lambda$. 
The $\Lambda$-Cold-Dark-Matter ($\Lambda$CDM) model 
\cite{peebles1982Largescale, peebles1984Tests} describes the standard cosmic evolution. 
However, there are some unsettled problems in the concordance model. 
One is the cosmological constant problem \cite{weinberg1989Cosmological,carroll2001Cosmological}. 
The measured energy density of $\Lambda$ is too small compared to the theoretical prediction. 
Another is the coincidence problem, which is why the present values of energy densities of 
$\Lambda$ and matter field have the same order of magnitude. 
Moreover, there exists non-negligible disagreement for the present value of 
matter fluctuation amplitude $\sigma_8$ between low- and high-redshift 
measurements \cite{macaulay2013Lower, hildebrandt2017KiDS450, nesseris2017Tension, joudaki2018KiDS450, Nunes:2021ipq}. There is also the similar tension for the today's 
Hubble constant $H_0$ \cite{riess2016DETERMINATION, verde2019Tensions, riess2019Large, freedman2019CarnegieChicago, reid2019Improved, wong2020H0LiCOW, shajib2020STRIDES, aghanim2020Planck}. 
For the high-redshift measurement, the values of these quantities are obtained from 
the observation of Cosmic-Microwave-Background by assuming the $\Lambda$CDM scenario. 
These problems may indicate the necessity of a new cosmological scenario 
other than the $\Lambda$CDM model. 

If the cosmological constant is not the origin of DE, then we need to find an alternative 
for explaining the late-time cosmic acceleration. In this context, the dynamical 
dark energy models based on scalar-tensor theories are well studied 
(see Refs.~\cite{Copeland:2006wr,Clifton:2011jh,Joyce:2014kja,kase2019Dark} for reviews). 
For scalar-tensor theories with a single scalar field, there is a framework 
which can treat the several theories in a unified manner,  
the so-called Horndeski theory \cite{horndeski1974Secondorder, deffayet2011Kessence, kobayashi2011Generalized, charmousis2012General} 
which is the most general scalar-tensor theory with second-order equations of motion. 
After the gravitational wave event GW170817 \cite{theligoscientificcollaboration2017GW170817} 
with its electromagnetic counterpart GRB170817A \cite{goldstein2017Ordinary}, 
the tensor propagation speed $c_t$ is tightly constrained to be quite close to 
that of light $c$. If we demand that $c_t=c$ in Horndeski theories, then the resultant theory 
uncoupled to CDM predicts the enhancement of the growth of matter perturbations 
relevant to the scale of galaxy clusterings. As a consequence, the aforementioned 
$\sigma_8$ tension tends to be even worse with respect to the $\Lambda$CDM model. 
 
In recent years, it is suggested that an interaction between DE and CDM can 
alleviate the $\sigma_8$ tensions 
\cite{zimdahl2001Interacting, wang2005Transition, amendola2007Consequences, guo2007Probing, wei2007Observational, valiviita2008Largescale, gavela2009Dark, salvatelli2014Indications, kumar2016Probing, divalentino2017Can, kumar2017Echo, an2018Relieving, kazantzidis2018Evolution, yang2018Largescale, yang2018Tale, pan2019Interacting, cardenas2020Interacting, divalentino2020Interacting, divalentino2020Nonminimal, jimenez2020Cosmological, vagnozzi2020We, yang2020Dark, divalentino2021Interacting, kang2021Reconstructing, kumar2021Remedy, peracaula2021Running, salzano2021JPAS, sinha2021Differentiating, yang20212021, yang2021Theoretical, Vagnozzi:2021quy}. 
Such a coupling has been introduced at first in order to resolve the coincidence problem 
\cite{amendola2000Coupled,Amendola:1999dr}. In the phenomenological approaches 
\cite{dalal2001Testing, chimento2003Interacting, pourtsidou2013Models, bohmer2015Interacting, bohmer2015Interactinga, skordis2015Parametrized, pourtsidou2016Reconciling, linton2018Variable, jesus2020Can, chamings2020Understanding, carrilho2021Interacting, figueruelo2021JPAS}, 
the interaction terms between CDM and scalar field $\phi$ associated with DE are 
added to their background equations of motion by hand. In this case, however, 
there is ambiguity in the way of defining perturbative quantities \cite{valiviita2008Largescale, valiviita2010Observational, tamanini2015Phenomenological}. 
In order to overcome the difficulties, the Lagrangian formulation of interactions is proposed 
in Refs.~\cite{koivisto2015ScalarFluid, dutta2017scalarfluid, kase2020Scalarfield, amendola2020Scaling, kase2020Weak, kase2020General, linton2021Momentum, jimenez2021Probing, jimenez2021Velocitydependent} in which 
CDM is described as a perfect fluid by the so-called Schutz-Sorkin action 
\cite{schutz1977Variational, brown1993Action,DeFelice:2009bx}. 
The interaction between DE and CDM consists of energy transfer and momentum exchange.  
The former is characterized by a CDM number density $n_c$ coupled to scalar field $\phi$ 
and its kinetic term $X = - \partial_{\mu}\phi \partial^{\mu}\phi / 2$, while the latter 
is implemented with a scalar product of the CDM four-velocity $u_c^{\mu}$ and 
field derivative $\partial_{\mu} \phi$, i.e., $Z = u_c^{\mu} \partial_{\mu} \phi$. 
In Ref.~\cite{kase2020General}, the cosmological perturbations are investigated 
with the general coupling Lagrangian $f(n_c, \phi, X, Z)$. It was shown that the coupling must have 
the form $f(n_c, \phi, X, Z) = - f_1(\phi, X, Z) \rho_c(n_c) + f_2(\phi, X, Z)$ under 
a condition for CDM velocity to vanish, where  $\rho_c$ is 
the CDM energy density. Between these two types of interactions, 
it is known that momentum exchange can suppress the growth of CDM density contrast \cite{pourtsidou2013Models, bohmer2015Interacting, bohmer2015Interactinga, skordis2015Parametrized, koivisto2015ScalarFluid, pourtsidou2016Reconciling, dutta2017scalarfluid, linton2018Variable, kase2020Scalarfield, chamings2020Understanding, amendola2020Scaling, kase2020Weak, Ferlito:2022mok} and may alleviate the $\sigma_8$ tension. 
However, since actual observations are executed targeting baryon in galaxies, 
it is difficult to conclude that the $\sigma_8$ tension can be completely relaxed 
as long as the scalar field couples only to CDM. Furthermore, in the context of 
compactified higher-dimensional theories, for instance, a scalar field generally 
couples to ordinary matter \cite{calcagni2005Dark, gumjudpai2005Coupled, carroll1998Quintessence}. 
It is therefore more natural to take into account the scalar field coupled not only to CDM 
but also to baryon.

On the other hand, the local gravity experiments in the solar system 
agree with general relativity in high precision~\cite{will2014Confrontation}. 
Thus, in scalar-tensor theories, the propagation of scalar field coupled to 
baryon (the so-called fifth force) must be screened on solar-system scales. 
One of several such screening scenarios is the Vainshtein mechanism 
\cite{vainshtein1972Problem, deffayet2002Accelerated, deffayet2002Nonperturbative, nicolis2004Classical, koyama2007Nonlinear, nicolis2009Galileon, koyama2011Analytic, babichev2013Introduction} in which derivative self-interaction of a scalar degree of freedom 
suppresses the propagation of fifth force at short distances.  The Vainshtein screening 
effect in Horndeski theories with the uncoupled matter is studied in Refs.~\cite{DeFelice:2011th,kimura2012Vainshtein,Kase:2013uja}. 
The analysis includes the case for subclass of Horndeski theories satisfying $c_t=c$, 
and shows that the Newton gravity can be recovered on sufficient small scales 
by virtue of the self-interaction term of scalar field. 

In this paper, we study the local gravity behavior of matter-coupled Horndeski theories 
with the general interaction Lagrangian $f(n, \phi, X, Z)$. The subclass of Horndeski 
theories satisfying $c_t=c$ is adopted in the DE sector. The matter sector is treated as 
a pressureless perfect fluid, which is described by a Schutz-Sorkin action. 
In the static and spherically symmetric spacetime adopted in 
Refs.~\cite{DeFelice:2011th,Kase:2013uja}, the momentum exchange 
$Z=u^{\mu} \partial_{\mu} \phi$ trivially vanishes at the background level 
since the four velocity of matter field $u^{\mu}$ is time-like while the scalar field is space-like. 
Alternatively, we consider the spherically symmetric perturbations on the cosmological 
background as in Ref.~\cite{kimura2012Vainshtein}, such that the momentum exchange 
survives even in the background. 
The scalar perturbation equations in the Newtonian gauge are obtained 
under the quasi-static approximation inside the sub-Hubble scales following 
the methods suggested in Refs.~\cite{kimura2012Vainshtein,Kobayashi:2014ida,langlois2018Scalartensor}. 
By solving these equations, we derive general formula of the Vainshtein 
radius and show that the screening effect successfully works even 
in the presence of a scalar field $\phi$ nonminimally and kinetically 
coupled to ordinary matter field. We also discuss the constraints on the coupling 
not to change the Vainshtein radius significantly.

This paper is organized as follows. In Sec.~\ref{sec2}, the background equations of motion are 
derived on the flat Friedmann-Lema\^{i}tre-Robertson-Walker space-time. 
In Sec.~\ref{sec3}, we obtain the scalar perturbation equations under the quasi-static 
approximation inside the sub-Hubble scales. We then explore the solutions to linearized 
perturbation equations. In Sec.~\ref{sec4}, we will find small scale solutions where 
the nonlinear terms for scalar field have dominant contributions in perturbation equations. 
We determine the Vainshtein radius and show the deviation of gravitational potentials from GR 
is screened deep inside the Vainshtein radius even in the presence of the coupling. 
The bounds on coupling terms are derived for the concrete models. 
Sec.~\ref{sec5} is devoted to conclusions.

We use the natural unit where the speed of light $c$, the reduced Planck constant $\hbar$, 
and the Boltzmann constant $k_B$, are equivalent to 1. 

\section{Horndeski theories with the matter coupling}
\label{sec2}

%
\subsection{Total action describing the Horndeski theories coupled to nonrelativistic matter}

We consider the subclass of Horndeski theories coupled to the matter field 
described by the following action: 
\begin{align}
\mathcal{S} 
=& 
\int {\rm d}^4 x \sqrt{-g} \, \mathcal{L}_{H}
- \int {\rm d}^4 x \left[ \sqrt{-g} \, \rho(n) + J^{\mu} \partial_{\mu} \ell \right]
\notag\\
&+\int {\rm d}^4 x \sqrt{-g} \, f(n, \phi, X, Z)\,,
\label{eq:act}
\end{align}
where $g$ is the determinant of metric tensor $g_{\mu \nu}$, and 
\begin{align}
\mathcal{L}_{\rm H} = G_4(\phi) \, R + G_2(\phi, X) + G_3(\phi, X) \Box \phi\,,
\label{eq:Horndeski}
\end{align}
is the Lagrangian density describing the subclass of Horndeski theories 
in which the propagation speed of gravitational waves is strictly equal to 
that of light 
\cite{bellini2014Maximal, lombriser2016Breaking, ezquiaga2017Dark, creminelli2017Dark, sakstein2017Implications, baker2017Strong, amendola2018Fate, crisostomi2018Selfaccelerating, kase2018Dark, kase2019Dark}. 
Here, $G_{2}$ and $G_{3}$ are functions of scalar field $\phi$ and its kinetic term,
\begin{align}
X = -\frac{1}{2} g^{\mu \nu} \partial_{\mu} \phi \partial_{\nu} \phi\,, 
\end{align}
with the covariant operator $\nabla_{\mu}$ and the d'Alembertian 
$\Box = g^{\mu \nu} \nabla_{\mu} \nabla_{\nu}$, while $G_{4}$ 
depends only on $\phi$ and represents the coupling between 
the scalar field and the Ricci scalar $R$. For $G_4(\phi)\neq$ constant, 
this nonminimal coupling gives rise to the modification of gravity. 
In the presence of the nonlinear self-interaction term $G_3(\phi,X)\Box\phi$, 
the Vainshtein screening suppresses such modification at short distances 
\cite{DeFelice:2011th,kimura2012Vainshtein,Kase:2013uja}. 

The second integral of the right-hand-side of Eq.~(\ref{eq:act}) corresponds to 
the Schutz-Sorkin action \cite{schutz1977Variational, brown1993Action,DeFelice:2009bx} 
describing the perfect fluid. We treat this perfect fluid as general nonrelativistic 
matter including not only CDM but also baryon. In Eq.~(\ref{eq:act}), 
the energy density $\rho$ is a function of fluid number density $n$, 
and the vector field $J^{\mu}$ is related to the fluid four velocity $u^{\mu}$, as 
\begin{align}
u^{\mu} = \frac{J^{\mu}}{n \sqrt{-g}}\,. 
\label{eq:def-u}
\end{align}
The four velocity obeys the relation $u^{\mu} u_{\mu} = -1$, which leads to 
the expression of $n$ in terms of $J^{\mu}$ as 
\begin{align}
n = \sqrt{ \frac{g_{\mu \nu} J^{\mu}J^{\nu}}{g} }\,.  
\label{eq:def-n}
\end{align}
The scalar quantity $\ell$ is a Lagrange multiplier with the notation 
$\partial_{\mu} \ell = \partial \ell / \partial x^{\mu}$. This quantity is 
related to the conservation of particle number on the cosmological 
background as we will see later. 

The third integral of the right-hand-side of Eq.~(\ref{eq:act}) represents 
the interaction between the scalar field and matter. The function $f$ depends on 
$n, \phi, X,$ and the quantity 
\begin{align}
Z = u^{\mu} \nabla_{\mu} \phi\,,
\end{align}
characterizing momentum exchange. In Ref.~\cite{kase2020General}, 
it has been shown that the coupling $f$ generally affects the fluid 
sound speed. Such an effect causes an increase of pressure for the nonrelativistic 
matter which can prevent the successful structure formation. 
This situation can be avoided if the coupling $f$ satisfies the following condition: 
\begin{align}
f_{,nn} = 0\,,
\label{eq:f-constraints}
\end{align}
where the comma in subscripts represents a partial derivative with respect to 
the scalar quantity represented in the index, e.g., $f_{,n} = \partial f / \partial n$ 
and $f_{,nn} = \partial^2 f / \partial n^2$. Under this condition, the functional form 
of the coupling $f$ is restricted to 
\begin{align}
f(n, \phi, X, Z) = - f_1 (\phi, X, Z)\, n + f_2(\phi, X, Z)\,.
\end{align}
Since the energy density of nonrelativistic matter linearly depends on 
its number density, the above coupling can also be expressed as 
\begin{align}
f(n, \phi, X, Z) = - f_1 (\phi, X, Z)\, \rho(n) + f_2(\phi, X, Z)\,. 
\label{eq:f-concrete}
\end{align}
In this section, we keep to use the general functional form $f(n,\phi,X,Z)$ 
for the sake of generality. We will adopt Eqs.~(\ref{eq:f-constraints})-(\ref{eq:f-concrete}) 
when needed in the following section . 

\subsection{Cosmological perturbations and background equations of motion}

We study spherically symmetric perturbations on the cosmological background in the Newtonian gauge,
\begin{align}
{\rm d}s^2 
= 
&- (1 + 2 \alpha) {\rm d}t^2 
\notag\\
&+ a(t)^2 (1 + 2 \beta) \left( {\rm d}r^2 + r^{2} \left[ {\rm d}\theta^{2} 
+ \sin^{2}\theta \, {\rm d}\varphi^{2} \right] \right)\,,
\label{eq:line-element}
\end{align}
where $a(t)$ is the time-dependent scale factor, $r$ is comoving distance, 
$\theta$ is polar angle, and $\varphi$ is azimuthal angle. 
The scalar perturbations $\alpha, \beta$ depend on only $t$ and $r$ 
since we consider spherically symmetric overdensity. 
The scalar field is decomposed into the time-dependent background part 
$\bar{\phi}(t)$ and the perturbed part $\delta \phi$, as
\begin{align}
\phi(t,r) = \bar{\phi}(t) + \delta \phi(t, r)\,.  
\label{eq:dphi}
\end{align}
Similarly, we decompose the energy density of matter,
\begin{align}
\rho(t,r) = \bar{\rho}(t) + \delta \rho(t, r)\,,
\end{align}
as well as the temporal and spatial components of $J^{\mu}$, 
\begin{align}
&J^0 = \sqrt{-\bar{g}} \, \left[ \, \bar{n}(t) + \delta J(t, r) \right]\,,\notag\\
&J^r = \sqrt{-\bar{g}} \, a(t)^{-2} \partial_{r} \delta j(t, r)\,,\notag\\
&J^{\theta} = J^{\varphi} = 0\,,
\label{eq:def-djdJ}
\end{align}
where $\delta J$, $\delta j$ are scalar perturbations. 
We hereafter omit the bar. The matter velocity potential $v$ is defined by 
\begin{align}
u_{r} = - \partial_{r} v\,.  
\label{eq:def-v}
\end{align}
We can replace $\delta j$ to $v$ for linear perturbations since a substitution of 
Eqs.~(\ref{eq:def-djdJ}) and (\ref{eq:def-v}) into Eq.~(\ref{eq:def-u}) results 
in the following relation,
\begin{align}
\partial_{r} \delta j = - n\, \partial_{r} v\,. 
\label{eq:dj-to-v}
\end{align}
We henceforth discuss velocity potential using $v$ instead of $\delta j$. 
Expanding Eq.~(\ref{eq:def-n}) for the line element (\ref{eq:line-element}) 
with Eq.~(\ref{eq:def-djdJ}) up to linear order in perturbations, 
the perturbed number density of matter fluid $\delta n$ is given by
\begin{align}
\delta n = \delta J - 3 n \beta\,. 
\label{eq:delta-n}
\end{align}
Hence, $\delta J$ can be eliminated by using relation (\ref{eq:delta-n}) 
with $\delta \rho = \rho_{,n} \delta n$, as
\begin{align}
\delta J = \frac{\delta \rho}{\rho_{,n}} + 3 n \beta\,. 
\label{eq:dJ-to-drho}
\end{align}
Variation of the action (\ref{eq:act}) in terms of $J^{\mu}$, it follows that
\begin{align}
\partial_{\mu} \ell 
= (\rho_{, n} - f_{, n}) u_{\mu} + \frac{f_{, Z}}{n} ( \nabla_{\mu} \phi + Z u_{\mu} )\,.
\label{eq:Dmu-ell}
\end{align}
On using Eqs.~(\ref{eq:dphi}) and (\ref{eq:def-v}), the leading order terms of 
Eq.~(\ref{eq:Dmu-ell}) are given by
\begin{align}
\dot{\ell} = - (\rho_{, n} - f_{, n})\,,
\label{eq:D0-ell}
\end{align}
for the temporal component on the background, 
where a dot represents the derivative with respect to $t$, and
\begin{align}
\partial_r \ell 
= 
- (\rho_{, n} - f_{, n}) \partial_r v 
+ \frac{f_{, Z}}{n} (\partial_r \delta \phi - \dot{\phi}\, \partial_r v)\,,
\label{eq:Di-ell}
\end{align}
for the radial component at the linear perturbation level. 
Integrating Eq.~(\ref{eq:D0-ell}) for $t$ and Eq.~(\ref{eq:Di-ell}) with respect to $r$, 
we obtain
\begin{align}
\ell 
= 
&- \int^t \left[ \rho_{, n} (\tilde{t}) - f_{, n}(\tilde{t}) \right] {\rm d}\tilde{t}
- \left( \rho_{, n} - f_{, n} + \frac{\dot{\phi} f_{, Z}}{n} \right) v\notag\\
&+ \frac{f_{, Z}}{n} \delta \phi\,.
\label{eq:ell-linear}
\end{align}
Hence, the quantity $\ell$ can be expressed in terms of background 
quantities and other perturbation variables.  
Additionally, varying the action (\ref{eq:act}) with respect to $\ell$ 
results in the following constraint,
\begin{align}
\partial_{\mu} J^{\mu} = 0\,. 
\label{eq:J-constraint}
\end{align}
On using Eq.~(\ref{eq:def-djdJ}), the constraint (\ref{eq:J-constraint}) leads to 
$n a^3 =$ constant, which corresponds to the conservation of 
total particle number of matter fluid. This relation is equivalent to the continuity equation, 
\begin{align}
\dot{\rho} + 3 H (\rho + P) = 0\,,
\end{align}
where $H = \dot{a}/a$ is the Hubble-Lema\^{i}tre expansion rate, and
\begin{align}
P = n \rho_{,n} - \rho\,,
\end{align}
is the pressure of matter fluid. We can derive the background equations of motion 
by expanding the action (\ref{eq:act}) up to linear order in time-dependent perturbations 
$\alpha, \beta, \delta \phi$ and eliminating $\delta j$, $\delta J$, $\ell$ with using 
Eqs.~(\ref{eq:dj-to-v}), (\ref{eq:dJ-to-drho}), and (\ref{eq:ell-linear}). 
Then, it follows that 
\begin{align}
& 3 M_{\rm pl}^{2} H^{2} = \rho_{\rm DE} + \rho\,, 
\label{eq:FriedmannEq-1} \\
& M_{\rm pl}^{2} \left( 2 \dot{H} + 3H^{2} \right) = - P_{\rm DE} - P\,,  
\\
& \dot{\rho}_{\rm DE} + 3H (\rho_{\rm DE} + P_{\rm DE}) = 0\,,
\end{align}
where $M_{\rm pl}$ is the reduced Planck mass related to 
the Newton gravity constant $G$ as $M_{\rm pl}=(8\pi G)^{-1/2}$. 
The quantities $\rho_{\rm DE}$ and $P_{\rm DE}$ correspond to the energy density 
and the pressure of scalar field associated with dark energy defined by
\begin{align}
\rho_{\rm DE} 
= 
&-G_2 + \dot{\phi}^2 G_{2,X} + \dot{\phi}^2 (G_{3,\phi}-3H\dot{\phi}G_{3,X})\notag\\ 
&+ 3H^2 (M_{\rm pl}^2 - 2G_{4}) - 6H \dot{\phi} G_{4,\phi} \notag\\
&- f + \dot{\phi}^2 f_{,X} + \dot{\phi} f_{,Z}\,,
\label{def-rhoDE}
\\
P_{\rm DE} 
= 
&G_2 + \dot{\phi}^2 \left( G_{3,\phi}+\ddot{\phi}G_{3,X} \right)\notag\\ 
&+ \left( 2\dot{H}+3H^2 \right) \left( 2G_4 - M_{\rm pl}^2 \right) 
+ 2 \left( \ddot{\phi}+2H\dot{\phi} \right) G_{4,\phi} \notag\\
&+ 2 \dot{\phi}^2 G_{4,\phi \phi} 
+ f - n f_{,n}\,,
\label{eq:def-PDE}
\end{align}
respectively. 

\section{Perturbation equations and the solutions in the linear regime}
\label{sec3}

Let us derive the scalar perturbation equations and clarify 
how the matter coupling affects the gravitational potentials 
deep inside the Hubble radius. 

\subsection{Perturbation equations under the quasi-static approximation}

We expand the action (\ref{eq:act}) up to quadratic order in scalar perturbations. 
Variations of the expanded action with respect to $\alpha$, $\beta$, 
$\delta \phi$, $v$, $\delta \rho$, $\delta \ell$ lead to perturbation equations. 
In what follows, we adopt the same procedure taken in 
Refs.~\cite{kimura2012Vainshtein,Kobayashi:2014ida,langlois2018Scalartensor}. 
First, we adopt the following approximation to perturbation equations deep inside 
the Hubble radius, i.e., $a\,r \ll 1/H$, where $r$ is the typical distance\footnote{
In Fourier transformed cartesian coordinates, this distance corresponds to 
a wave length of fluctuation $1/k$ where $k$ is a wave number.}. 
For the quantity $\xi$ representing $\alpha$, $\beta$, or $\delta \phi$,
we ignore time derivatives $\dot{\xi}$ and non-derivative terms $\xi$ 
compared to spatial derivatives $\xi'/a$ since $|\xi'/a| \sim |\xi / (a\,r)|$, whereas the time derivative 
$|\dot{\xi}|$ takes the order of $|H \xi|$. We apply $|\xi'/a|^2 \ll |\xi'| / (a\,r)$ due to 
$|\xi'/a| \ll 1/(a\,r)$ while we keep the higher-order derivatives of $\delta \phi$ 
since the terms can grow up at short distances. 
Second, in order to focus on a static overdensity, we eliminate $t$-dependence 
in matter perturbations $\delta \rho$ and $v$. From the Poisson equation, 
we deal with $\delta \rho, v$ as comparable to spatial derivatives $\xi'/a$. 
The equations for $\delta \rho$ and $v$ correspond to the $t$ and $r$ 
components of Eq.~(\ref{eq:Dmu-ell}). The combination of these equations is equivalent to 
Eq.~(3.38) in Ref.~\cite{kase2020General}, which is needed to discuss the time evolution of 
$\delta \rho$. In the following, we do not use them, since we consider 
static overdensity, but focus on the equation for $\delta\ell$. 
After an integration\footnote{We set the boundary conditions that $\alpha', \beta', \delta \phi', v', \alpha, \beta, \delta \phi, v$ vanish when $\delta \rho$ is absent.} over $r$, this equation is written as,
\begin{align}
v' 
= \frac{6 a^2 H (1 + \dot{P}/\dot{\rho}) A \, r}{\rho+P}
= \frac{6 a^2 H A \, r}{\rho}\,,
\label{eq:Edl}
\end{align}
with
\begin{align}
A = \frac{M(r)}{8 \pi \bar{r}^{3}}, \qquad
M(r) = 4 \pi \int \bar{r}^{2} \delta \rho(\bar{r}) {\rm d}\bar{r}\,,
\end{align}
where $\bar{r}$ is the physical distance defined by $\bar{r}=ar$, and we used 
the conditions for dust-like matter: $\dot{P}/\dot{\rho} = 0$ and $P = 0$. 
We apply the relation (\ref{eq:Edl}) in order to eliminate the velocity potential $v$ from 
the other perturbation equations. In the discussion below, we use $\bar{r}$ instead of 
$r$ itself but omit the bar for the sake of simplicity. 

After the process, the reduced equations for 
$\alpha$, $\beta$, $\delta \phi$ can be integrated over $r$ once. We rewrite the perturbed 
quantities with the notation, $x = \delta \phi' / r$, $y = \alpha' / r$, and $z = \beta' / r$, 
following the same notation adopted in 
Refs.~\cite{kimura2012Vainshtein,Kobayashi:2014ida,langlois2018Scalartensor}. 
Consequently, we obtain 
\begin{align}
&
\frac{c_{1} x}{2 } + q_{t} z 
+ \left( 1 + \frac{c_{2} n}{\rho} \right) A  = 0\,,
\label{eq:IEalpha}
\\
&\frac{\dot{q}_{t} x}{ \dot{\phi}} 
+ q_{t} (y + z)
 + \frac{c_{3} n^{2} A}{2 \rho} = 0\,,
\label{eq:IEbeta}
\\
&
\frac{(c_{1} \dot{\phi}-\dot{q}_{t}) x^{2}}{\dot{\phi}^{3}} 
+ \frac{c_{4} x}{2} 
- \frac{c_{1} y}{2} 
- \frac{\dot{q}_{t} z}{\dot{\phi}}
\notag\\
&
+ \frac{\left[ (c_{5} n^{2} + c_{7}) H - c_{6} n \right] A}{2 \rho} = 0\,,
\label{eq:IEdphi}
\end{align}
where
\begin{align}
q_{t} =&\,\, 
2 G_{4}\,,
\label{eq:def-qt} \\
c_{1} =&\,\, 
G_{3,X} \dot{\phi}^{2} + 2 G_{4,\phi}\,,
\label{eq:def-c1} \\
c_{2} =&\,\, 
f_{,Xn} \dot{\phi}^{2} + f_{,Zn} \dot{\phi} - f_{,n}\,,
\label{eq:def-c2} \\
c_{3} =&\,\, 
6 f_{,nn}\,,
\label{eq:def-c3} \\
c_{4} =&\,\,  
G_{2,X} + 2 G_{3,\phi}  - 4 G_{3,X} \dot{\phi} H
 - 2 G_{3,X} \ddot{\phi}\notag\\
& - G_{3,X\phi} \dot{\phi}^{2}- \dot{\phi}^{2} \ddot{\phi} G_{3,XX}+ f_{,X}\,,
\label{eq:def-c4} \\
c_{5} =&\,\, 
6 (f_{,Xnn} \dot{\phi} + f_{,Znn})\,, 
\label{eq:def-c5}\\
c_{6} =&\,\, 
2 \left[(\ddot{\phi} f_{,XXn} + f_{,Xn\phi}) \dot{\phi}^{2} 
+ (f_{,Xn} + f_{,ZZn}) \ddot{\phi}
\right. \nonumber \\
&\left.
+ (3 H f_{,Xn} + 2 f_{,XZn} \ddot{\phi} + f_{,Zn\phi})\dot{\phi} 
+ 3 f_{,Zn} H - f_{,n\phi}
\right]\,, 
\label{eq:def-c6}\\
c_{7} =&\,\, 
6 f_{,Z}\,. 
\label{eq:def-c7}
\end{align}
Due to the conditions (\ref{eq:f-constraints}), we find 
\begin{align}
c_3 = c_5 = 0\,.
\end{align}
We eliminate them in the following discussion.

\subsection{The linear regime solutions in the presence of matter coupling}

At large distances from the origin of overdensity, the quantity $A$ is much 
smaller than unity so that the linear terms in Eq.~(\ref{eq:IEdphi}) should 
dominate over the quadratic term of $x$. As a result, all the Eqs.~(\ref{eq:IEalpha})-(\ref{eq:IEdphi}) settle down to be linear, then we can easily obtain the solution 
on large scales as
\begin{align}
&\alpha' = \left[
G_{N}\left(1+ \frac{c_{2} n}{\rho}\right) 
+ \frac{ C (\dot{\phi} c_{1} - 2 \dot{q}_{t}) }{16\pi H q_{t}^{2}} 
\right]  \frac{M}{r^{2}}\,,
\label{eq:solalpha-lin}
\\
&\beta' = -\left[
G_{N}\left(1+ \frac{c_{2} n}{\rho}\right) 
+ \frac{C \dot{\phi}c_{1} }{16 \pi H q_{t}^{2}}
\right]  \frac{M}{r^{2}}\,,
\label{eq:solbeta-lin}
\\
&\delta \phi' = \frac{C \dot{\phi}}{8 \pi H q_{t}} \frac{M}{r^{2}}\,,
\label{eq:soldphi-lin}
\end{align}
with
\begin{align}
&G_{N} =
\frac{1}{8 \pi q_{t}}\,,
\label{eq:def-Geff}
\\
&C =
\frac{2 q_{t} H}{\dot{\phi} \rho}
\frac{\left[ H c_{7} q_{t} - (c_{1} c_{2} + c_{6} q_{t}) n 
 - c_{1} \rho \right] \dot{\phi} + 2 \dot{q}_{t} (c_{2} n + \rho)}
{ (c_{1}^{2} - 2 c_{4} q_{t})\dot{\phi} - 4 \dot{q}_{t} c_{1}}\,,
\label{eq:def-C}
\end{align}
where $G_N$ corresponds to the gravitational coupling. 
Since the coefficients $c_2$, $c_6$, $c_7$ and the last term in the 
definition of $c_4$ arise from the coupling, the interaction between 
the scalar field and matter affects the gravitational potentials via 
the term $c_2n/\rho$ and the quantity $C$ in Eqs.~(\ref{eq:solalpha-lin})-(\ref{eq:solbeta-lin}). Without them, 
Eqs.~(\ref{eq:solalpha-lin})-(\ref{eq:soldphi-lin}) reduce to ones given by 
Eqs.~(33)-(35) in Ref.~\cite{kimura2012Vainshtein}. In this linear regime, 
the parametrized post-Newtonian (PPN) parameter, defined by 
$\gamma=-\beta/\alpha$, is given as 
\begin{align}
\gamma = \left[ 1 - \frac{2 C \dot{q}_{t}}
{16\pi G_{N}(1+nc_2/\rho) H q_{t}^{2}+C \dot{\phi} c_{1}} \right]^{-1}\,.
\label{eq:gamma}
\end{align}
The gravity in the linear regime is obviously modified when $G_{4,\phi} \neq 0$, while 
there seems no modification for the GR-like case characterized by $G_{4,\phi} = 0$ 
since $\gamma = 1$. 
However if we may consider the case with $f = -f_1(\phi) \rho(n)$ 
as a theory in Einstein frame, we will see that the modification of 
gravity exists even though $G_{4,\phi} = 0$. Let us consider 
the simple action in Einstein frame described by
\begin{align}
S_{\rm E} =& \int {\rm d}^4 x \sqrt{-\hat{g}} 
\left( \frac{M_{\rm pl}^2}{2} \hat{R} 
- \frac{1}{2}\hat{\nabla}_{\mu}\phi \hat{\nabla}^{\mu}\phi \right)\notag\\
&+ \hat{S}_m (\hat{\Psi}_{\rm m}, \hat{g}_{\mu \nu}, f_1(\phi))\,,
\label{eq:E-frame}
\end{align}
where a hat represents the quantity in Einstein frame, and 
$\hat{\Psi}_m$ is a matter field. If we conformally transform the metric 
as $g_{\mu \nu} = \Omega^{-2}(\phi) \, \hat{g}_{\mu \nu}$ such that 
the matter coupling vanishes after the transformation, i.e., the Jordan frame, 
the resultant action can be written as 
\begin{align}
S_{\rm J} =& \int {\rm d}^4 x \sqrt{-g} 
\left( \frac{M_{\rm pl}^2}{2}\Omega^2 R 
- \frac{1}{2} \nabla_{\mu} \phi \nabla^{\mu} \phi \right)
\notag\\
&
+ S_m(\Psi_m, g_{\mu \nu})\,.
\label{eq:J-frame}
\end{align}
The energy-momentum $\hat{T}^{\mu}_{\nu}$ derived from the action 
(\ref{eq:E-frame}) and $T^{\mu}_{\nu}$ from (\ref{eq:J-frame}) are 
connected to each other as 
\be
\hat{T}^{\mu}_{\nu} = \Omega^{-4} T^{\mu}_{\nu}\,. 
\label{T1}
\ee
Meanwhile, for the case $f = -f_1(\phi) \rho(n)$, the coupled energy-momentum 
tensor $\hat{T}^{\mu}_{\nu}$ is expressed by (see, for instance, Eq.~(2.27) 
in Ref.~\cite{kase2020General})
\begin{align}
\hat{T}^{\mu}_{\nu}= f_1 T^{\mu}_{\nu} \,.
\label{T2}
\end{align}
Comparing Eqs.~(\ref{T1})-(\ref{T2}), we obtain 
\be
\Omega^2=f_1^{-1/2}\,.
\label{Om}
\ee
The resultant Jordan frame action (\ref{eq:J-frame}) with (\ref{Om}) is included in 
Eq.~(\ref{eq:act}) with $G_4=(M_{\rm pl}^2/2)f_1^{-1/2}$ and $f=0$. Hence, 
$\gamma\neq1$ holds in Jordan frame. This can also be interpreted as follows. 
When we decompose the conformal factor as 
$\Omega(\bar{\phi}+\delta \phi) = \bar{\Omega}(\bar{\phi}) 
(1 + \delta \Omega(\bar{\phi}, \delta \phi))$, 
the gravitational potential in Jordan frame are given by 
$\alpha = \hat{\alpha} - \delta \Omega$. We also obtain 
$\beta$ in the same way, and they lead to $\gamma\neq1$. 
Hence, $f_1(\phi)$ affects the gravitational law even for $G_{4,\phi} = 0$. 
In a similar manner, $f_2$ can also change the linear regime gravity. 
Thus, we need the screening mechanism at short distances 
in the presence of the coupling. 

\section{Screening mechanism in the nonlinear regime}
\label{sec4}

%
\subsection{The nonlinear regime solutions in the presence of matter coupling}

Let us proceed to the analysis at short distances where the nonlinear term of 
scalar field in Eq.~(\ref{eq:IEdphi}) becomes non-negligible. On using 
Eqs.~(\ref{eq:IEalpha})-(\ref{eq:IEbeta}) to eliminate $y$ and $z$ from 
Eq.~(\ref{eq:IEdphi}), the resultant equation includes only $x$ and $A$. 
Rewriting them to $\delta \phi$ and $M$, it reduces to 
\begin{align}
\frac{B}{2 H \dot{\phi} \, r} \delta \phi'^2 + \delta \phi'
- \frac{\dot{\phi}\,C M}{8 \pi q_t H r^2} = 0\,,
\label{eq:eqdphi}
\end{align}
where the quantity $C$ is given by Eq.~(\ref{eq:def-C}), and 
\begin{align}
B = \frac{8 H q_{t}}{\dot{\phi}}
\frac{\dot{q}_{t} - \dot{\phi}c_{1}}
{\dot{\phi} c_{1}^{2} - 2 \dot{\phi} q_{t} c_{4}- 4 \dot{q}_{t} c_{1}}\,.
\label{eq:def-B}
\end{align}
Solving the above equation for $\delta\phi'$, we obtain 
\begin{align}
\delta \phi' = 
\frac{H \dot{\phi} \,r}{B} \left( \sqrt{1 + \frac{B C M}{4 \pi H^{2}q_{t} r^{3}}} - 1 \right)\,.
\label{eq:soldphi}
\end{align}
Here, we took the branch in which the solution is consistent with 
the linear regime solution (\ref{eq:soldphi-lin}) in the large scale 
limit $r\to\infty$. The above solution also reduces to 
Eq.~(\ref{eq:soldphi-lin}) in the limit $B\to0$ which corresponds to 
$G_{3,X}\to 0$ since $\dot{q_t}-\dot{\phi}c_1=-\dot{\phi}^3G_{3,X}$ 
in Eq.~(\ref{eq:def-B}). The vanishing of self-interaction term 
$G_3$ leads to the absence of nonlinear term in Eq.~(\ref{eq:eqdphi}), 
so that Eqs.~(\ref{eq:soldphi-lin}) and (\ref{eq:soldphi}) coincide with each other. 
We consider the case with $G_{3,X}\neq0$ in what follows. 
As is clear from Eq.~(\ref{eq:soldphi}), the second term in the square root 
of Eq.~(\ref{eq:soldphi}) gets larger on smaller scales and exceeds unity inside 
the radius $r_V$ given by 
\begin{align}
r_{V} = \left( \frac{B C M}{4\pi H^2 q_{t}} \right)^{1/3}\,.
\label{eq:def-rV}
\end{align}
This is the so-called Vainshtein radius. 
Under the condition $r\ll r_V$, Eq.~(\ref{eq:soldphi}) reduces to 
\begin{align}
\delta \phi' \simeq 
\frac{H \dot{\phi} \, r_V}{B} \left( \frac{r_{V}}{r} \right)^{1/2}
\,,
\label{eq:soldphi-screened}
\end{align}
which shows that the solution deep inside the Vainshtein radius, $\delta\phi'\propto r^{-1/2}$, 
varies more slowly than the linear regime solution, $\delta\phi'\propto r^{-2}$. The qualitatively 
different behavior of $\delta\phi$ in the region $r\ll r_V$ compared to the linear regime enables 
the Vainshtein mechanism at work. Indeed, substituting Eq.~(\ref{eq:soldphi-screened}) into 
Eqs.~(\ref{eq:IEalpha})-(\ref{eq:IEbeta}) and solving them in terms of the derivatives of 
gravitational potentials, $\alpha'$ and $\beta'$, we obtain 
\begin{align}
&\alpha' = 
\frac{G_{N}M}{r^{2}}
\left[1+ \frac{c_{2} n}{\rho}
+\frac{C(\dot{\phi}c_1-2\dot{q}_t)}{Hq_t}\left(\frac{r}{r_V}\right)^{3/2}
\right]\,,
\label{eq:solalpha-screened}
\\
&\beta' = 
-\frac{G_{N}M}{r^{2}}
\left[1+ \frac{c_{2} n}{\rho}
+\frac{C\dot{\phi}c_1}{Hq_t}\left(\frac{r}{r_V}\right)^{3/2}
\right]\,. 
\label{eq:solbeta-screened}
\end{align}
The solutions (\ref{eq:soldphi-screened})-(\ref{eq:solbeta-screened}) settle down to 
Eqs.~(42), (44), and (45) in Ref.~\cite{kimura2012Vainshtein}, when the coupling vanishes.
Eqs.~(\ref{eq:solalpha-screened})-(\ref{eq:solbeta-screened}) are integrated to give,  
\begin{align}
&\alpha = 
-\frac{G_{N}M}{r}
\left[1+ \frac{c_{2} n}{\rho}
-\frac{2C(\dot{\phi}c_1-2\dot{q}_t)}{Hq_t}\left(\frac{r}{r_V}\right)^{3/2}
\right]\,,
\label{solal}
\\
&\beta = 
\frac{G_{N}M}{r^{2}}
\left[1+ \frac{c_{2} n}{\rho}
-\frac{2C\dot{\phi}c_1}{Hq_t}\left(\frac{r}{r_V}\right)^{3/2}
\right]\,.
\label{solbe}
\end{align}
Obviously, the terms proportional to $(r/r_V)^{3/2}$ in the solutions 
of gravitational potentials, as well as the derivative of scalar field 
perturbation given by Eq.~(\ref{eq:soldphi-screened}), 
become negligible deep inside the Vainshtein radius, $r\ll r_V$. While 
the term $c_2n/\rho$ sourced by the matter coupling remains even 
in this regime, it can be absorbed into the definition of $M$ both in 
Eqs.~(\ref{solal})-(\ref{solbe}) such as $\tilde{M}=(1+c_2n/\rho)M$. 
Indeed, the PPN parameter reduces to $\gamma\simeq1$
in the regime $r\ll r_V$. Consequently, we have found that the Vainshtein 
screening efficiently works even in the presence of the coupling 
$f(n, \phi, X, Z)$.

\subsection{Bounds on the coupling terms in concrete theories}

We now investigate a bound on the coupling $f_{1}$ and $f_{2}$ 
in the context of gravity at short distances. 
In order to put detailed bounds on these functions, we need to specify 
the functional forms of them together with the Horndeski functions 
$G_2$, $G_3$, $G_4$, and study the background evolution in the model 
so as to determine the present value of background quantities, e.g., 
$\phi$, $\dot{\phi}$, $H$. However, this is beyond the scope of our paper. 
Alternatively, we put rough bounds on the functions $f_{1}$ and $f_{2}$ 
by requiring that the coupling terms do not drastically change the Vainshtein radius. 
In doing so, we focus on the theories described by $G_{2} = X$ and 
$G_{3} = \kappa_{3} X$, where $\kappa_{3}$ is a constant, for the sake of simplicity. 
First we consider the pure momentum exchange, 
i.e., $f_{1} = 0$ and $f_{2,Z}\neq0$. In this case, the Vainshtein radius defined by 
Eq.~(\ref{eq:def-rV}) reduces to 
\begin{align}
r_{V}^{3} = 
&\,\,
(8 \kappa_{3} G_{4} M/\pi \rho)
\left[ (\kappa_{3} \dot{\phi}^{2} - 2 G_{4,\phi}) \rho 
- 12 H G_{4} f_{2, Z} \right] /
\notag\\
 &\left\{ \kappa_{3}^{2} \dot{\phi}^{4} + 4 \left[ 2 (2 \dot{\phi} H + \ddot{\phi}) G_{4} 
 - G_{4,\phi}\dot{\phi}^{2} \right] \kappa_{3} 
\right.
\notag\\
&
- 4 (1 + f_{2,X}) G_{4} - 12 G_{4,\phi}^{2}\Big\}^{2}.
\end{align} 
In order for the momentum exchange $f_{2}$ not to significantly modify 
the Vainshtein radius, the coupling should satisfy the following two conditions: 
\begin{align}
&\mathcal{O}(f_{2,X}) \lesssim \mathcal{O}(1)\,, 
\label{eq:f2-bound-1} \\
&\mathcal{O}\left(\frac{f_{2,Z}}{M_{\rm pl}^{2}}\right) \lesssim 
\mathcal{O}\left( \frac{(\kappa_{3} \dot{\phi}^{2} - 2G_{4,\phi}) H}{G_{4}} \times 10^{-1} \right), \label{eq:f2-bound-2}
\end{align}
where we used the observational bounds on the matter density parameter 
at present epoch, $\Omega_{m, 0}\simeq0.3$, such that 
\begin{align}
\mathcal{O}\left( \frac{\rho}{12 M_{\rm pl}^{2} H^{2}} \right)
= \mathcal{O}\left(\frac{\Omega_{m,0}}{4}\right)
\simeq \mathcal{O}(10^{-1})\,.
\end{align}

Second, we consider the pure energy transfer, i.e., $f_{1} = f_{1}(\phi)$ and $f_{2} = 0$. 
In this case, the Vainshtein radius reduces to
\begin{align}
r_{V}^{3} = &\,\,
(8 \kappa_{3}G_{4} M/\pi)\left[ (\kappa_{3}\dot{\phi}^{2} - 2 G_{4,\phi})(1+f_{1}) + 4 G_{4} f_{1,\phi} \right] /
\notag\\
&
\left\{ \kappa_{3}^{2} \dot{\phi}^{4} + 4 \left[ 2(2\dot{\phi} H + \ddot{\phi}) G_{4} - G_{4,\phi} \dot{\phi}^{2} \right] \kappa_{3} 
\right.\notag\\
&
- 4 G_{4} - 12 G_{4,\phi}^{2} \Big\}^{2} \,.
\end{align}
In order for $f_1$ not to modify the Vainshtein radius drastically, 
the bound on $f_{1}$ is given as
\begin{align}	
\mathcal{O}(f_{1}) \lesssim \mathcal{O}(1)\,, 
\label{eq:f1-bound}
\end{align}
and $f_1$ must not be a steep function with respect to $\phi$.

\section{Conclusions}
\label{sec5}

We studied the Vainshtein screening of coupled DE and matter theories. 
In Sec.~\ref{sec2}, we introduced the total action (\ref{eq:act}) describing such theories. 
For the DE sector, we adopted the subclass of Horndeski theories (\ref{eq:Horndeski}) 
in which the speed of gravitational waves is equal to that of light. The matter was dealt with 
as a pressureless perfect fluid which is described by the Schutz-Sorkin action. 
These two components are coupled to each other through the general interacting Lagrangian 
$f(n, \phi, X, Z) = - f_1 (\phi, X, Z) \, \rho(n) + f_2(\phi, X, Z)$ satisfying the condition 
(\ref{eq:f-constraints}). We adopted the perturbed metric with the Newtonian gauge 
(\ref{eq:line-element}) to derive the equations of motion 
(\ref{eq:FriedmannEq-1})-(\ref{eq:def-PDE}) on the cosmological background.

In Sec.~\ref{sec3}, we obtained the perturbation equations (\ref{eq:IEalpha})-(\ref{eq:IEdphi}) 
for the spherically symmetric and static configurations on top of the cosmological background. 
We adopted the quasi-static approximation deep inside the Hubble scale following the methods 
suggested in Refs.~\cite{kimura2012Vainshtein,Kobayashi:2014ida,langlois2018Scalartensor}. 
In this approximation scheme, time- and non-derivative terms of $\alpha$, $\beta$, $\delta \phi$ 
were neglected compared to their spatial derivatives. The perturbations $\delta \rho$, $v$ 
of the perfect fluid were considered as static, to focus on the local gravity on the solar system scales. 
By solving the perturbation equations (\ref{eq:IEalpha})-(\ref{eq:IEdphi}) in the linear regime 
where the quadratic term of $x$ becomes negligible, we found the solutions 
(\ref{eq:solalpha-lin})-(\ref{eq:soldphi-lin}) which leads the PPN parameter $\gamma$ 
given by Eq.~(\ref{eq:gamma}). The PPN parameter reduces to $\gamma = 1$ when 
$G_{4,\phi} = 0$ holds. However, as long as the coupling exists, $\gamma$ can deviate 
from unity through a conformal transformation (see also Ref.~\cite{hui2009Equivalence}). 
Hence, the screening mechanism needs to be at work on small scales.

In Sec.~\ref{sec4}, we derived the closed form equation (\ref{eq:eqdphi}) 
for the perturbation of scalar field, and determined the Vainshtein radius 
$r_V$ in Eq.~(\ref{eq:def-rV}) inside which the nonlinear term gives 
the dominant contribution. 
Deep inside the Vainshtein radius, i.e., $r \ll r_V$, the solutions of 
a perturbed scalar field and two gravitational potentials reduce to those 
given in Eqs.~(\ref{eq:soldphi-screened})-(\ref{eq:solbeta-screened}). 
We showed that the modification of gravity depends on $r$ in the form of 
$(r/r_V)^{3/2}$ being negligible in the region $r\ll r_V$, and confirmed that 
the PPN parameter $\gamma$ reduced to unity in this regime. 
Thus, we have found that the Vainshtein mechanism sufficiently works 
even in the presence of the coupling $f(n, \phi, X, Z) = - f_1 (\phi, X, Z) \, \rho(n) + f_2(\phi, X, Z)$.
Furthermore, by requiring that the coupling does not drastically change the Vainshtein radius, 
we put rough bounds on the functions $f_1$ and $f_2$ in concrete theories. 
For the pure momentum exchange characterized by $f_1=0$ and $f_{2,Z}\neq0$, 
we specified the two bounds on $f_2$, Eqs.~(\ref{eq:f2-bound-1})-(\ref{eq:f2-bound-2}). 
We also derived the similar bound (\ref{eq:f1-bound}) on $f_1$ in the pure energy 
transfer case satisfying $f_1=f_1(\phi)$ and $f_2=0$. In order to make an accurate 
estimate of the Vainshtein radius and put more precise constraints on $f_1$ and $f_2$ 
from the local gravity experiments, we need to specify a model and solve 
the background equations (\ref{eq:FriedmannEq-1})-(\ref{eq:def-PDE}) numerically. 
It would be interesting to clarify how the coupling $f$ decrease/increase the $r_V$ 
in the viable interacting models, but the discussion is beyond the scope of this paper.

We note that, in the viable coupled dark energy models such as those studied in Refs.~\cite{amendola2020Scaling,kase2020Weak,jaman2022What}, a field potential plays key role for the late-time cosmic acceleration. In the case where the field potential gives the dominant contributions even at short distances, one can consider another type of screening scenario, the chameleon mechanism \cite{khoury2004Chameleon,khoury2004Chameleona}. One can also consider the case in which both the chameleon potential and the non-linear kinetic term associated with Vainshtein mechanism coexist \cite{Ali:2012cv}. Then, it would be of interest to study whether the chameleon mechanism can also work sufficiently even in the presence of the general coupling $f(n,\phi,X,Z)$.

In this paper, we focused on quasi-static perturbations in small scales paying attention 
to the gravitational experiments within the solar system. In doing so, we neglected 
the time-dependence in the matter density contrast. However, the DE-matter interaction 
affects the evolution of density fluctuation in galaxy cluster scales. It would be of interest 
to put constraint on the coupling from Large Scale Structure observations. 

\section*{Acknowledgements}

RK is supported by the Grant-in-Aid for Young Scientists 
of the JSPS No.~20K14471. 


\bibliography{grqc}
\bibliographystyle{grqc.bst}

\renewcommand{\appendix}{}
\appendix
\renewcommand{\thesection}{Appendix \Alph{section}}
\renewcommand{\thesubsection}{\Alph{section}.\arabic{subsection}}
\setcounter{section}{0}
\setcounter{subsection}{0}


\end{document}